# ЗАЩИТА НА ДОКУМЕНТИ, ПУБЛИКУВАНИ В ИНТЕРНЕТ, ОТ НЕПРАВОМЕРЕН ВЪНШЕН ДОСТЪП

Йордан Калмуков

# PROTECTING ONLINE DOCUMENTS FROM AN UNAUTHORIZED EXTERNAL ACCESS

**Yordan Kalmukov**

***Abstract:*** *The modern multi-tier web applications and information systems store and process various types of data. Some of them are stored in a database, controlled by an external database management system, while other data are stored directly within the server's file system. The database is secured by the database management system itself, but it is a programmer's responsibility to design and develop a security protection of the files managed by the information system.*

*This paper summarizes the existing and suggests new rules for design and implementation of an in-depth security protection of file resources, published on the Internet, from an unauthorized external access.*

## *Въведение*

Появата на Интернет в края на XX век поставя начало на нов етап в развитието на компютърните технологии. Появяват се разнообразни приложни програми, позволяващи на човек да използва компютрите не само за работа, но и за комуникация и забавление. Това от своя страна дава допълнителен тласък в бързото развитие на глобалната мрежа, превръщайки я в неразделна част от живота на хората.

Осъзнали силата на тази нова медия, частните компании се надпреварват да предлагат на потребителите си разнообразни (често дори и ненужни) услуги, борейки се да привлекат все повече и по-платежоспособни рекламодатели. Настоящата година доказа, че Интернет необратимо промени не само бизнеса, но и световната политика. Благодарение на web-базираните социални мрежи „гражданите" на иначе силно консервативните и религиозни арабски диктатури успяха да се обединят в желанието си за повече свобода и справедливост, и дори да свалят от власт диктатори, управлявали от десетилетия.

На фона на всичко това през последните години се засили тенденцията голяма част от традиционните десктоп приложения да се проектират и реализират като web приложения и в последствие да се предлагат дори не като софтуер, а като услуги. Web приложенията, за разлика от традиционните десктоп приложенията, не се инсталират локално на машината на потребителя, а на сървър. Те са типичен пример за клиент-сървър архитектура, при която множество физически отдалечени потребители използват едно и също приложение, инсталирано на сървъра [1]. Именно отдалеченият достъп определя най-важните им предимства пред традиционните приложения – възможността за достъп до приложението и/или данните, които управлява то, по всяко време на денонощието и от всяка точка на света. Единственото необходимо на потребителя за да използва дадено web приложение е връзка с Интернет и Интернет браузър. Понастоящем браузъри има вградени във всяка една операционна система, дори и в мобилните телефони.

Съвременните многослойни web приложения и информационни системи (ИС) оперират с множество и различни по тип данни. Част от тях се управляват от външни за приложението системи за управление на бази от данни (СУБД), а други се запазват на сървъра директно във вид на файлове (документи, изображения и т.н.). Независимо къде и как точно се съхраняват данните потребителският достъп до тях трябва да се извършва единствено и само през конкретната информационна система, която ги управлява. Директният достъп от външни приложения и/или лица трябва да бъде невъзможен или ограничен само до потребители с административни, за сървъра, права. Защитата и управлението на достъпа до данните, съхранявани от СУБД се поема от самата СУБД, но задължение и отговорност на програмиста и системния администратор е да осигурят защитата и контрола на достъп до данните съхранени директно във файловата система на сървъра.

Тази статия обобщава най-важните правила и предлага едно допълнително, с решаваща роля (правило 3), за защита на файлови ресурси, публикувани в Интернет, от неправомерен директен достъп от страна на външни потребители. Предложените подходи са особено полезни в случаите на споделен хостинг, когато програмистът няма нужните права за управление на web сървъра и предоставената му част от файловата система е изцяло достъпна по HTTP протокола.

### *Как и къде да съхраняваме данните – в база от данни, управлявана от СУБД или директно във файловата система?*

Тъй като проектирането и програмирането са много вариантни, то еднозначен отговор на този въпрос няма. Зависи какъв е типът на данните, колко често ще се обновяват и какво ниво на защита е необходимо.

Релационните БД и системите за управление на бази от данни намират огромно приложение в web програмирането и стоят в основата на всяка информационна система и динамичен сайт. Голямата им популярност се дължи преди всичко на това, че СУБД предоставя на програмиста:

- **Управление на едновременен многопотребителски достъп**. Ако данните не се управляват от СУБД, то тогава програмистът трябва да предвиди съответните механизми и да реализира необходимите възможности приложението само да управлява кой във всеки един момент може да чете или да редактира данните; да укаже какъв е приоритетът на операциите и какво се случва когато в един и същ момент даден потребител се опитва да прочете данните, а друг да ги редактира. Управлението на конкурентния (едновременния) многопотребителски достъп не е никак лесна задача. Поемането й от СУБД дава възможност на програмиста да мисли и проектира приложението си на по-високо ниво и да не се занимава с тривиални, макар и сложни, но вече решени проблеми.

- **_Ефективен и бърз достъп до информацията (търсене)_**. Системите за управление на бази от данни се проектират и реализират от множество експерти, работещи години наред за получаването на един наистина добър краен резултат. Колко и да се опитва даден програмист няма как сам и за кратко време да измисли по-добър и ефективен алгоритъм за търсене от този вграден в която и да е СУБД.

От казаното до тук става ясно, че всички данни, подлежащи на честа актуализация в комбинация с едновременен много потребителски достъп, както и тези по които често ще се търси следва да се съхраняват в БД. Достъпът до съвременните релационни СУБД се извършва почти изцяло с помощта на специализирания език от свръх високо ниво SQL чрез потребителски SQL заявки. С увеличаването на обема на базата от данни и степента на фрагментиране на данните, времето за изпълнение на SQL заявките и достъп до данните също се увеличава. Именно затова никак не е добра идеята всички данни да се управляват от СУБД, просто защото е по-лесно.

Къде и как точно ще се съхраняват данните зависи преди всичко от типа им както следва:

- *Текстови и променливи във времето данни*, описващи обекти и/или процеси се управляват от СУБД. Такива са примерно данните за потребители, продукти, поръчки, взаимовръзки между обекти, мета данни за документи и т.н.
- *Текстови данни, които не следва да се променят* (от потребителите) по време на жизнения цикъл на приложението е по-добре да се изнесат във външни файлове. Такива са имената на държавите, статуси на обекти и процеси и т.н. Те се задават твърдо от разработчиците и не могат да бъдат променяни от потребителите.
- *Двоични данни и статични ресурси* като изображения, музика, видео, документи се съхраняват директно във файловата система. Търсенето по тях обикновено се извършва на базата на предварително снети мета данни, например заглавие, изпълнител, цветови характеристики, резюмета и т.н. Тези мета данни в общия случай се съхраняват в БД, управлявана от СУБД, като по този начин се постига ефективно търсене и на ресурсите съхранени във вид на файлове. Тези ресурси са наречени статични, защото информационната система, която ги управлява не може по никакъв начин да ги модифицира. Тя просто управлява достъпа до тях и нищо повече.

Управлението на достъпа и защитата на всички данни, съхранявани директно във файловата система на сървъра е грижа и отговорност на програмиста и системния администратор.

### *Ролята на сървъра и системните администратори*

Web приложенията се инсталират на web сървър. Сървърът може да бъде *частен, предназначен само за конкретното приложение* (на английски dedicated server) или *споделен* (на английски shared server/hosting). Споделеният сървър от своя страна може да бъде *частен споделен* или *публичен споделен*. В случая на частен споделен сървър няколко приложения споделят един и същ сървър, но тези приложения са собственост или се администрират от една и съща фирма и съответно до сървъра имат достъп само служители на фирмата, но не външни лица. При публичния споделен хостинг всеки потребител, който реши може за малка сума да си закупи правото да използва дадения сървър, като споделя ресурсите му с множество (обикновено стотици) други потребители.

За съжаление да се постигне високо ниво на сигурност в среда на споделен хостинг никак не е лесно. Причината е в това, че за да се изпълни даден сорс код или да се изпрати даден статичен ресурс към клиента, web сървърът трябва да има права да го прочете от файловата система [2]. При споделен web сървър това означава, че приложенията на всички потребители имат достъп до всички файлове, до които има достъп самият web сървър, т.е. до файловете и на останалите потребители. В условията на публичен споделен хост защитата на данните може да се разглежда в два аспекта: *защита от потребителите, споделящи същия сървър*, която тема е детайлно разгледана в литературата (примерно в книгата на Крис Шифлет „Основи на PHP сигурността" [2]); и *защита от външни злонамерени потребители*, което е и целта на тази статия. Разбира се в условията на частен сървър остава само необходимостта от защита от външни потребители.

### *Правила за защита на файлови ресурси от неправомерен външен достъп*

Независимо дали частен или публичен споделен е сървърът, той се конфигурира, така че част от файловата му система да бъде видима през Интернет (фигура 1), като достъпът до видимите ресурси се извършва посредством комуникационния протокол HTTP. За различните web (HTTP) сървъри конфигурирането става по различен начин, но основната идея е същата. Примерите в тази статия използват конфигурационни директиви и се отнасят за Apache HTTP Server.

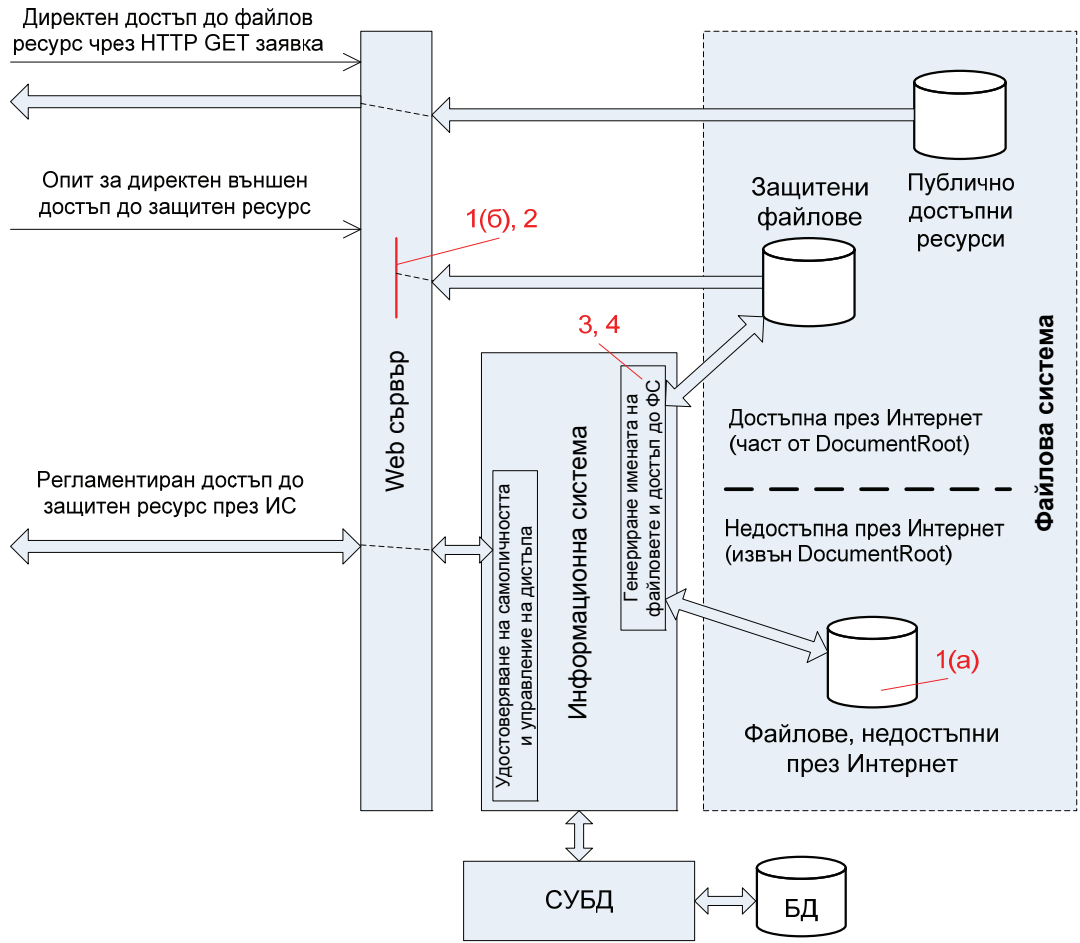

*Фиг. 1. Обобщена архитектура, отразяваща взаимодействието на информационната система с web сървъра, системата за управление на бази от данни и локалната файлова система. Числата в червено показват местата, на които се реализират защитите, базирани на предложените правила от 1 до 4.*

Основното изискване към сигурността на всяка информационна система е данните и файловете, които управлява тя, да бъдат достъпни единствено и само през нея. Достъпът на външни лица и / или приложения трябва да бъде невъзможен. Директно следствие от това изискване са правилата 1(а) и 1 (б).

**_Правило 1(а)_**: Документите (файловите ресурси), които ще се защитават, се съхраняват в тази част на файловата система, която не е видима през Интернет, т.е. по HTTP протокола. Документите се прочитат през файловата система и в последствие приложението изпраща прочетената информация до потребителя. Последният няма директен достъп до файловете.

При Apache web server, директивата `DocumentRoot` указва кореннатадиректория на т.нар. web дърво. Нейното съдържание и съдържанието на всички нейни поддиректории е видимо в Интернет. Желателно е защитените документи да се разполагат извън тази директория. Най-често обаче, дори и при частен споделен сървър, на потребителите им се дава достъп само до файловата система, видима през Интернет. Тогава реализацията на правило 1(а) е невъзможна и се преминава към следващата алтернатива – 1(б).

***Правило 1(б):*** Документите (файловите ресурси), които ще се защитават, се съхраняват в поддиректория на `DocumentRoot`, т.е. видима през Интернет, но web сървърът се настройва така, че да забрани достъпа на всички потребители до тази директория и нейното съдържание. Отново документите са достъпни единствено през файловата система и приложението след като ги прочете ги изпраща на потребителя. Последният няма директен достъп до файловете.

За забрана на достъпа до дадена директория, web сървърът най-често се настройва динамично посредством специален конфигурационен файл, наречен (в термините на Apache) .htaccess файл. За целта в него се разполагат следните директиви:

```
Order Deny, Allow
Deny from all
```

Директивата „`Deny from all`" забранява достъпа до дадената директория по HTTP, включително и от локалната машина. Настройките важат рекурсивно само за директорията, в която се намира конфигурационният файл. В много случаи динамичната забрана на достъп е разрешена дори при публичен споделен сървър, т.е. реализацията на правило 1(б) често (не винаги) е реалистична. В случай, че се реализира защита по правило 1(а) или 1(б), то реализацията и на правило 4 е задължителна, защото то дава единствената възможност за достъп до защитените ресурси, тъй като директният достъп чрез HTTP GET заявка е забранен.

Ако по една или друга причина програмистът / администраторът на приложението няма права да конфигурира сървъра, то правилата 1(а) и 1(б) са неприложими. В този случай съдържанието на директорията с документите остава видимо през Интернет и отговорност на програмиста е да приложи допълнителни механизми, с които да скрие доколкото е възможно информацията. В зависимост от настройките на web сървъра е твърде вероятно той да генерира листинг на съдържанието на съответната директория. Ако това се случи злонамереният външен Интернет потребител безпроблемно придобива достъп както до списъка на файловете, така и до самите файлове в съответната директория. За да се избегне това е необходимо комбинирано да се използват правила 2 и 3.

***Правило 2:*** Създава се индексен HTML файл, наименуван `index.html`, който се разполага в директорията, в която се намират и защитените документи. Файлът може да бъде празен или да съдържа съобщение от типа „Достъпът до тази директория е забранен". Поради наличието на индексен файл web сървърът, независимо от настройките си, няма да генерира списък на файловете в директорията, а ще изведе съдържанието на файла. Въпреки, че реално достъпът до дадената директория не е забранен (поради пропадането на защити 1(а) и 1(б)) потребителят може да стигне до защитените файлове само ако ги адресира директно. За целта той трябва да отгатне имената им, което не винаги е лесно или възможно.

В случай на масово използвани приложения с отворен код, които се разпространяват безплатно, отгатването на имената на защитените ресурси е сравнително лесно, ако не са взети други мерки. Web-базираните информационни системи обикновено преименуват файловете си преди да ги съхранят. Целта на това е от една страна да се избегне възможността за дублиране на имената, и от друга да се улесни идентификацията им и достъпа до тях в последствие. Тук ключова роля играе системата / „алгоритъма" на наименуване. Ако то е от типа от 1 до N .<разширение на файла> е очевидно, че имената могат да бъдат лесно отгатнати, особено в случаите когато става въпрос за приложения, чиито програмен код по една или друга причина е достъпен за външните потребители.

За да се намали риска от отгатване на имената и директно адресиране на ресурсите е необходимо да се реализира защита според правило номер 3.

*Правило 3:* Защитените ресурси трябва да бъдат преименувани, така че злонамерените потребители да не могат да отгатнат имената им и да ги адресират директно. Най-добре е в имената на пръв поглед да няма никаква логика, но все пак ИС трябва да може да ги генерира за да осъществява достъп до тях. Един добър подход е имената да бъдат md5 хешове [3], получени от данни, до които злонамереният потребител няма достъп.

Ако защитените ресурси са .pdf документи, предавани от потребителите то техните имена могат да се получат от хеширането на конкатенацията от:
- потребителско име;
- дата (времеви маркер) на предаване / качване на документа;
- таен ключ, който се съхранява само на сървъра.

Нека потребителят Иван Иванов с потребителско име iivanov качва някакъв pdf документ. Тайният ключ на сървъра е „azsymbabameca". Нека времевият маркер да бъде в UNIX формат, т.е. броя секунди от 1 Януари 1970, примерно: 1306090530. Тогава името на файла ще се генерира така:
```
md5("iivanov"."1306090530"."azsymbabameca");
```
и в конкретния случай ще бъде: `693076a03195395ed5215a3ac0d3e70e.pdf`

Когато имената на файловете се състоят от 32 шестнадесетични символа от типа:
```
6bd36deecd3332838d4c55e456994bc5.pdf
b58de5617deb7ee8a9396de12d83b784.pdf
```
злонамереният потребител практически не би могъл да ги отгатне, нито да ги генерира, защото дори и от някъде да разбере времевия маркер на предаване на файла няма да знае тайния ключ, който се съхранява само на сървъра. Ако имената на файловете станат по-дълги дори и brute-force атаки не биха помогнали за налучкването.

Доказателства за значимостта и ролята на правило 3 може да се намерят навсякъде в реалността. Примерно информационната система за управление на научни конференции The MyReview System [4]. Това е една от най-добрите системи в дадената област и въпреки това проблеми със сигурността не липсват. При нея (версия 1.x) е реализирана защита само по правило 2, а защитените ресурси (научните доклади) се съхраняват в директория /FILES/ (видима през Интернет) с имена 1.pdf, 2.pdf и т.н. по реда на тяхното постъпване. Тъй като системата се разпространява безплатно и с отворен код е изключително лесно злонамерените потребители да видят как се казва директорията, в която са докладите, и че те се наименуват по указания начин. Защитата по правило 2 само предотвратява генерирането на списък на съдържанието на защитената директория, но не блокира достъпа. След като злосторниците знаят и името на директорията, и имена на защитените файлове, тогава могат да ги адресират директно и съответно web сървърът ще им ги изпрати. По този начин всеки злонамерен потребител може да се сдобие с още непубликувана информация и ако иска да я публикува от свое име. Организаторите на конференцията в този случай носят наказателна отговорност за неправомерно разгласяване на поверителна информация или безстопанственост към защитени с авторски права данни.

Правилата 2 и 3 са най-ефективни ако към тях се приложи и следващото правило 4.

*Правило 4:* Информационната система не трябва в никакъв случай да предоставя директен достъп чрез HTTP GET заявки до защитените документи, дори и на потребителите, имащи съответните права. Т.е. защитеният ресурс не трябва да се третира от web сървъра като статичен, а вместо това ИС трябва да го прочете от файловата система, да генерира съответните HTTP хедъри за отговор и да изпрати прочетените данни на клиента.

Т.е. достъпът до даден, иначе статичен, ресурс вместо да се осъществи директно чрез хипер връзка (HTTP GET заявка) от типа:
`<a href="/docs/resurs.pdf">Изтегли защитения ресурс</a>`
трябва да стане по указания от правилото начин.
Следва примерна реализация на PHP.

```
$d="/srv/www/testApp/docs/b58de5617deb7ee8a9396de12d83b784.pdf";
if ($fp=@fopen($d, "r")) {
    header("Content-Type: application/pdf");
    header("Content-Length: ".filesize($d));
    header("Accept-Ranges: bytes");
    header("Content-Disposition: attachment;
                        filename=\"yourFile.pdf\"");
    fpassthru($fp); exit();
}
```

`$d` е адресът на защитения ресурс (в случая pdf файл). Чрез функцията `fopen()` се отваря ресурса за четене. Ако прочитането е успешно

приложението изпраща няколко HTTP хедъра към клиента, с които указва какви са типът и размерът на следващите данни, а хедърът `Content-Disposition: attachment;` [5] принуждава потребителския браузър да изведе диалоговия прозорец за запазване на входящите данни във вид на файл или директното им изпращане към външно приложение на клиентската машина. Функцията `fpassthru()` изпраща прочетеното от файла `$d` съдържание към клиента.

Ако е реализирана защита по едно от правилата 1(а) или 1(б), то тогава достъпът до ресурсите трябва да става според правило 4 – друга алтернатива няма, тъй като ресурсите са недостъпни чрез директни HTTP заявки. Ако защитата е организирана на база правила 2 и 3 тогава не е задължително, но е силно препоръчително достъпът да се извърши отново според правило 4. Изобщо липсата на директни връзки до защитените ресурси гарантира, че достъпът до тях ще се осъществява винаги посредством информационната система, която ги управлява.

Посочените правила от 1 до 4 помагат да се блокира или затрудни до максимална степен директният достъп на *външни, за информационната система, лица до защитените ресурси*. Те обаче са безсилни в случаите, когато злонамерен потребител успее неправомерно да се сдобие с реални *чужди* права за достъп до информационната система. Затова, за да имат смисъл тези правила е необходимо и самата информационна система да предлага достатъчно високо ниво на сигурност и да не позволява външни потребители да използват неправомерно функционални възможности, предназначени само за регистрирани такива. Разработващите ИС трябва да предвидят и реализират всички необходими защити срещу прихващане на потребителка информация, най-вече потребителски имена и пароли; атаки от тип *фиксиране* и *кражба* на потребителски сесии (*session fixation* и *session hijacking*); неправомерно придобиване на права чрез атаки от типа *SQL injection* и др.

Организирането на защитата в дълбочина на практика означава да се реализират всички защити според посочените правила, така че ако някоя от тях пропадне поради технически причини или бъде пробита да сработят останалите.

*Заключение*

Тенденцията все по-голяма част от информационните системи да се реализират като web-базирани води до силно завишаване на изискванията за сигурност и защита на данните, тъй като „пренасянето" им в Интернет ги прави от затворен тип системи в достъпни практически за всеки злонамерен потребител по света. Основното изискване към всяка ИС, а именно данните, които управлява да бъдат достъпни единствено и само през нея, е по-трудно постижимо в Интернет среда. Въпреки, че много експерти съветват защитените данни да се съхраняват само в БД, управлявана от СУБД, особено в условията на споделен хостинг, това в редица случаи е нецелесъобразно и води до значително влошаване на производителността на приложението. Необходимо е да се вземе мотивирано решение и да се намери балансът между това кои данни ще бъдат съхранявани в БД и кои директно във файловата система.

Предложените в тази статия правила за организиране на защита в дълбочина спомагат напълно да се елиминира (в някои случаи) или изключително да се затрудни неправомерния директен достъп до защитените файлови ресурси. В случай, че не е възможно достъпът до защитените ресурси да бъде блокиран чрез настройване на web сървъра или чрез разполагането на ресурсите извън видимата част на файловата система, тогава е необходимо да се реализират задължително и трите правила – 2, 3 и 4 – само тогава защитата ще бъде ефективна. Ако едно от тези правила се пропусне съществува голяма вероятност да се пробият защитите, базирани на останалите две правила. Реализацията на 1(а) и 1(б) до голяма степен зависи от правата на програмиста (администратора на приложението) да настройва web сървъра, докато правила 2, 3 и 4 не са технически зависи от сървъра изобщо и могат да бъдат реализирани във всички случай. Отговорност на програмиста е да предвиди и реализира тези защити.

Предложените правила за организиране на защита в дълбочина са реализирани, тествани и успешно приложени в системата за управление на научни конференции, използвана от CompSysTech и ADBIS 2007. Реализираните защити в нея са тествани от докторанти в университета в Санкт Петербург, Русия.